\documentclass[12pt,preprint]{aastex}
\usepackage{psfig,natbib,amsmath}

\shorttitle{VLBI of GRB\,030329}
\shortauthors{Taylor et al.}
\begin{document}

\def\cit{2}
\def\vla{1}
\def\grb{GRB\,030329}
\def\simlt{\mathrel{\hbox{\rlap{\hbox{\lower4pt\hbox{$\sim$}}}\hbox{$<$}}}}
\def\simgt{\mathrel{\hbox{\rlap{\hbox{\lower4pt\hbox{$\sim$}}}\hbox{$>$}}}}

\title{~~\\ ~~\\ The Angular Size and Proper Motion of the Afterglow of
  GRB\,030329}

\author{G. B. Taylor \altaffilmark{\vla}, D. A.
Frail\altaffilmark{\vla}, E. Berger\altaffilmark{\cit} \&\ S. R.
Kulkarni\altaffilmark{\cit} }

\altaffiltext{\vla}{National Radio Astronomy Observatory, Socorro,
        NM 87801; gtaylor@nrao.edu, dfrail@nrao.edu}

\altaffiltext{\cit}{Division of Physics, Mathematics and Astronomy,
        105-24, California Institute of Technology, Pasadena, CA
        91125}

\begin{abstract}

The bright, nearby (z=0.1685) gamma-ray burst of 29 March 2003 has
presented us with the first opportunity to directly image the
expansion of a GRB.  This burst reached flux density levels at
centimeter wavelengths more than 50 times brighter than any previously
studied event. Here we present the results of a VLBI campaign using
the VLBA, VLA, Green Bank, Effelsberg, Arecibo, and Westerbork
telescopes that resolves the radio afterglow of \grb\ and constrains
its rate of expansion.  The size of the afterglow is found to be
$\sim$0.07 mas (0.2 pc) 25 days after the burst, and 0.17 mas (0.5 pc)
83 days after the burst, indicating an average velocity of 3-5$c$.
This expansion is consistent with expectations of the standard
fireball model.  We measure the projected proper motion of \grb\ in
the sky to $<$0.3 mas in the 80 days following the burst.  In
observations taken 52 days after the burst we detect an additional
compact component at a distance from the main component of 0.28 $\pm$
0.05 mas (0.80 pc).  The presence of this component is not expected
from the standard model.

\end{abstract}

\slugcomment{As Accepted by the Astrophysical Journal Letters}


%\vfill\eject
\section{Introduction}

The fireball model has provided a remarkably successful framework in
which to interpret observations of gamma-ray bursts (GRBs) and their
afterglows (see M\'esz\'aros 2002\nocite{mes02} for a recent review).
Among the earliest observational tests of this model was the
superluminal motion inferred for GRB\,970508 \citep{fkn+97}. Large
fluctuations in the radio flux density on timescales of several hours
were attributed to diffractive scattering of the radio waves as they
propagated through the turbulent ionized gas of our galaxy
\citep{goo97}. The observed ``quenching'' of this scintillation
pattern at $t\sim$2 weeks lead to a determination of a size of 3
$\mu$as, implying a mean apparent motion of 4$c$ which
was consistent with estimates from an expanding shock
\citep{fkn+97,wkf98}.  The motion appears superluminal due to a
geometric effect first described by \cite{ree66}.  This size
measurement was independently supported by constraints derived from
late-time calorimetry of GRB\,970508 \citep{fwk00}.

While interstellar scintillation can be a powerful tool to test
predictions of GRB afterglow models, such observations are rare.  Most
GRBs are expected to remain in the strong diffractive regime for only
a few days \citep{goo97,wal97}, and hence there are practical
difficulties obtaining well-sampled light curves. Nearby GRBs
($z<0.2$), although equally rare, enable us to image the afterglow
{\it directly}, measuring the expansion or proper motion of the
emitting region \citep{wax97b,sar99,gps99b,gl03}. Previous
applications of the technique of Very Long Baseline Interferometry
(VLBI) did not yield constraining limits on the angular size for 
distant GRBs ($z\sim1$) \citep{tfbk97,tbfk98}. This
situation changed with the detection of \grb\ by the {\it HETE-II}
spacecraft \citep{vcd+03}. At $z=0.1685$ \citep{gpe+03} \grb\ is one
of the nearest GRBs detected to date, and also the brightest. At
centimeter wavelengths it reached peak flux densities of 50 mJy,
motivating us to undertake a comprehensive multi-epoch VLBI
program. In this paper we describe the results of the first five
epochs covering the time between 3 and 83 days after the burst.

\section{Observations and Results}\label{sec:obs}

The five epochs are summarized in Table \ref{Observations}. All
observations employed the Very Long Baseline Array (VLBA) of the
NRAO\footnote{The National Radio Astronomy
Observatory is operated by Associated Universities, Inc., under
cooperative agreement with the National Science Foundation.}.  Other
telescopes used in one or more epochs include the Effelsberg 100-m
telescope\footnote{The 100-m telescope at Effelsberg is operated by
the Max-Planck-Institut f{\"u}r Radioastronomie in Bonn.}, the phased
VLA, the Green Bank Telescope (GBT), the Arecibo telescope, and the
Westerbork (WSRT) tied array.  The observing runs were typically 5
hours long, with 256 Mbps recording in full polarization with 2 bit
sampling.  The nearby (1.5$^\circ$) source J1051+2119 was used for
phase-referencing with a 2:1 minute cycle on source:calibrator below
15 GHz and a 1.3:1 minute cycle on source:calibrator at 15 GHz and
above.  The weak calibrator J1048+2115 was observed hourly to check on
the quality of the phase referencing.  Self-calibration was used to
further refine the calibration and remove atmospheric phase errors
when the signal-to-noise was sufficient.  
% save space
% This was generally possible
% except for the early VLBA-only observations and the 22 GHz observation
% on May 19.

From our 8.4 GHz observations on April 6 we derive a 3$\sigma$ limit
on the linear polarization of 0.16 mJy/beam, which corresponds to a
limit on the fractional polarization of $<$1.0\%.  In a
contemporaneous optical observation \cite{gre03} measure a
polarization of 2.2+/-0.3\%.  The decrease in polarization at lower
frequencies could be explained as the result of the source being
optically thick at 8.4 GHz at these early times.

For each observation listed in Table \ref{Observations} we fit a
circular Gaussian to the measured visibilities to derive angular
diameters (or limits).  For components that are slightly resolved, a
Gaussian, uniform disk, and ring all have a similar quadratic
dependence on baseline length for the short baselines \citep{tjp99}.
There are some differences in scaling in
that a Gaussian with size 1 mas FWHM is equivalent to
a uniform disk with diameter 1.6 mas, or to a ring with diameter 1.1 mas.  The
results of the Gaussian fits are summarized in Table \ref{Big_Gaussian}. Note
that sizes smaller than the synthesized beam are measurable because of
the high signal-to-noise. Significant source sizes of 70 $\mu$as and
170 $\mu$as are measured on April 22 and June 20, respectively.  
% save space
% This requires that \grb\ expanded with an average apparent velocity
% of $5c$ and $3c$, respectively. 
Uncertainties in the size estimates
were derived from the signal-to-noise ratios and the synthesised beam.
As a further check, we performed 100 Monte-Carlo simulations of the
data using identical ($u,v$) coverage, similar noise properties, and a
Gaussian component of known size added.  We found the variance of the
recovered sizes, by modelfitting in the same way we treat the
observations, to be a factor of two lower than the estimated 1
$\sigma$ uncertainties shown in Table \ref{Observations}.

The Gaussian fits made to the measured visibility data as a function
of baseline length returned residuals consistent with noise in all but
one case. The fit to the 15 GHz observation on May 19 produces a
significant residual ($>$20 sigma) which is $\sim$30\% of the peak
flux density and offset to the northeast at 0.28 $\pm$ 0.05 mas from
the main component. The exact nature of this second component
is not known but it would require an average velocity of 19$c$ to
reach its offset from the flux centroid. This component is not
detected in the less sensitive 22 GHz image taken on the same day, nor
was it seen in the fifth epoch at 8.4 GHz.

In addition to angular size, a best fit position was derived for each
observation in Table \ref{Observations}. These positions are listed in
Table \ref{Big_Gaussian}, expressed for convenience as offsets from
the 8.6 GHz position on April 1, and plotted in Fig.~1.  Error bars
were estimated from the variation in the position of \grb\ over the
course of the observation, added in quadrature with an estimated
systematic uncertainty in the position of 0.2 mas.  This systematic
term arises from imperfect modeling of the atmosphere and is roughly
proportional to the angular separation between the phase-reference
calibrator and the target source. 
% save space
% Solving for proper motion using all
% the data, we derive $\mu_{\rm r.a.}=-0.02\pm 0.80$ mas yr$^{-1}$ and
% $\mu_{\rm dec.}=-0.44\pm 0.63$ mas yr$^{-1}$. 
Note that the two
observations at 5 GHz suffer from a large, systematic offset, most
likely due to the increased contribution of the ionosphere at this
frequency.  The phase check source, J1048+2115, exhibited a similar
systematic offset at 5 GHz.  Discarding the measurements at 5 GHz, and
solving for proper motion, we derive $\mu_{\rm r.a.}=-0.32\pm 0.58$
mas yr$^{-1}$ and $\mu_{\rm dec.}=-0.31\pm 0.66$ mas yr$^{-1}$, or an
angular displacement over the first 80 days of 0.10$\pm$0.14 mas.

\section{Constraints on GRB Afterglow Models}
\subsection{Angular Size Measurements}\label{sec:size}

A gamma-ray burst drives a relativistic blast wave into a circumburst
medium of density $\rho$ whose radius $R$ is related to the energy of
the explosion approximately by $E\sim R^3\rho c^2\gamma^2$, where
$\gamma$ is the bulk Lorentz factor of the fireball. A more precise
description of the dynamics is given by the Blandford \& McKee
(1976)\nocite{bm76} solution. The {\it apparent} radius $R_\perp$ of
the relativistic blast wave as seen by a distant observer viewing a
GRB close to face on is approximately $R_\perp\sim R/\gamma$. A
calculation by Galama et al.~(2003)\nocite{gfs+03} for constant
density gives,

\begin{equation}\label{eqn:ism}
  R_\perp(ISM) = 3.9 \times
  10^{16}\left(\frac{E_{52}}{n_\circ}\right)^{1/8}
  \left(\frac{t_d}{1+z}\right)^{5/8}~{\rm cm}
\end{equation}
where $E_{52}$ is the isotropic energy normalized to $10^{52}$ erg and
$t_d$ is the time in days in the observers frame. The coefficient in
Eqn.~\ref{eqn:ism} is the same as in \cite{gps99b} but it is 6\%
larger than the estimate by \citep{wkf98}. A circumburst medium shaped
by mass loss from a massive progenitor star is expected to have a
density that falls off with radius as $\rho=A R^{-2}$, where
$A=\dot{M_w}/4\pi V_w$ is a constant, typically normalized to
A=5$\times 10^{11}A_*$ g cm$^{-1}$ ({\em i.e.,} values for the mass
loss rate $\dot{M_w}$ and wind velocity $V_w$ of a typical WR star).
Chevalier \& Li (2000)\nocite{cl00} give estimates for the
line-of-sight radius $R$ and Lorentz factor in a wind-blown medium but
for consistency we use Galama et al.~(2003)\nocite{gfs+03}:

\begin{equation}\label{eqn:wind}
  R_\perp(Wind) = 2.4 \times
  10^{16}\left(\frac{E_{52}}{A_*}\right)^{1/4}
  \left(\frac{t_d}{1+z}\right)^{3/4}~{\rm cm}
\end{equation}
Assuming a Lambda cosmology with $H_0 = 71$~km/s/Mpc, $\Omega_M =
0.27$ and $\Omega_\Lambda=0.73$, the angular-diameter distance of
\grb\ at $z=0.1685$ is $d_A=589\,$Mpc. Thus the angular {\it
  diameters} corresponding to the radii in Eqns.\ref{eqn:ism} and
\ref{eqn:wind} are $8.9(E_{52}/n_\circ)^{1/8}(t_d/1+z)^{5/8}$ $\mu$as
and $5.5(E_{52}/A_*)^{1/4}(t_d/1+z)^{3/4}$ $\mu$as for the ISM
and Wind models, respectively.  

The measured sizes 24 and 83 days after the burst give average
apparent perpendicular expansion velocities of 5 and 3$c$
respectively.  Based on energetics and breaks in the light curves,
typical GRBs appear to be collimated into a cone of angle, $\theta \sim
0.1$ radians \citep{ber03}, which we must be within to see the
gamma-rays.  Since the apparent expansion, $\beta_\perp$, is given
by $\beta_\perp = \beta \sin\theta /(1-\beta\cos\theta)$, then to get an
apparent superluminal expansion of 5$c$ requires Lorentz factors of
$\sim$7, and values of $\beta$ close to unity.  These values for
$\beta$ are much larger than $\sim$0.1 found in simulations
by \cite{can04}.

%We have assumed so far that the blast wave is isotropic, but in
%general GRB outflows are collimated into jets. Jets modify the
%dynamics since the radius of the fireball stays roughly constant and
%sideways expansion dominates \citep{rho99,sph99}, independent whether
%the medium is wind-blown or has a constant density. In this case the apparent
%radius $R_\perp\propto t^{1/2}$ which grows after the jet break time
%$t_j$ approximately as
%$R_\perp(JET)$=$R_\perp(t_j)\times(t/t_j)^{1/2}$, where $R_\perp(t_j)$
%is Eqn.~\ref{eqn:ism} (or Eqn.~\ref{eqn:wind}) evaluated at $t_d=t_j$.
%
%In time the expanding jet slows to non-relativistic velocities as it
%approaches spherical symmetry. The dynamics are then described by the
%Sedov-Taylor-von Neumann solution in which the radius in a constant
%density medium is given by $R=\zeta (E t^2/n_\circ m_p (1+z)^2)$,
%where $\zeta$ is an order unity factor that depends on the gas
%adiabatic index \citep{fwk00}. The timescale $t_{NR}$ for this
%transition is defined when $\dot{R}/c=1$. We take the estimates
%directly from Eqns. 3 and 9 of Livio \& Waxman (2000)\nocite{lw00}
%which are $t_{NR}=38(E_{52}/n_\circ)^{1/4}t_j^{1/4}$ and
%$t_{NR}=22(E_{52}/A_*)^{1/2}t_j^{1/2}$ for the ISM and wind models,
%respectively. After $t_{NR}$ the expansion slows considerably for the
%ISM model with $R_\perp\propto{t}^{2/5}$, while
%$R_\perp\propto{t}^{2/3}$ for a wind.

In Fig.~\ref{fig:theta} we show the evolution of the expected angular
size of \grb\ for the ISM and Wind models for some representative
values of $E_{52}/n_\circ$ (or $E_{52}/A_*$) and jet break times.
There is good agreement from the predictions of the isotropic models
with our measurements (Table \ref{Big_Gaussian}). The size estimates
in the ISM model are more robust than the wind model in the sense that
they are relatively insensitive to the ratio of $E_{52}/n_\circ$,
yielding values of 60 $\mu$as and 130 $\mu$as for $E_{52}/n_\circ\sim
1$ at $\Delta t$=25 and $\Delta t$=83 days, respectively.

The introduction of a jet raises the energy requirements
substantially. Part of this increase may be due to our limited
understanding of how to describe the lateral expansion of a GRB jet.
However, for $t_j\sim10$ d an acceptable fit is obtained for
$E_{52}/n_\circ\simeq$10 (or $E_{52}/A_*\simeq$7) and a best fit for
$E_{52}/n_\circ\simeq$30. For $t_j\sim0.5$ d acceptable fits for either
the ISM or wind models require $E_{52}/n_\circ\simgt$100.

There exists a wide range of energy estimates for \grb\ in the
literature.  Vanderspek et al.~(2004)\nocite{vsb+04} estimate an
isotropic gamma-ray energy release E$_{\gamma,iso}=(1.80\pm
0.07)\times 10^{52}$ erg. Allowing for a reasonable radiative
efficiency $\eta_\gamma$=0.2 \citep{pk01b},
$E_{52}=E_{\gamma,iso}/\eta_\gamma$=9. A jet-break seen at $\sim$0.5 d
in the X-ray and optical light curves \citep{pfk+03,tmg+03} reduces
this energy by a factor about 400. However, \cite{gnp03} in explaining
the unusual fluctuations in the optical light curve, increase the
energy by an factor of 10 by having the afterglow shock ``refreshed''
by slower moving ejecta shells. 

%Berger et al.~(2003)\nocite{bkp+03}
%(see also Lipkin et al.~2004\nocite{log+04}) propose a two-component
%jet model in which the bulk of the radio afterglow emission from \grb\
%originates from a wide-angle jet ($\theta_j=17^\circ$) with
%$t_j\sim$10 d and $E_{52}/n_\circ\simeq 0.3$. This choice of
%parameters predicts an angular size on day 83 of just 0.09 mas, about
%2$\sigma$ below our measurement.  The single-jet model of
%\cite{woo+04} has $t_j\sim$0.5 d and $E_{52}/n_\circ\simeq 15$.

\subsection{Proper Motion Limits}\label{sec:proper}

In the relativistic fireball model a shift in the flux centroid is
expected due to the spreading of the jet ejecta \citep{sar99}. For a
jet viewed off the main axis the shift can be substantial
\citep{gl03}. However, since gamma-rays were detected from \grb\ it is
likely that we are viewing the jet largely on axis.  The predicted
displacement in this case is expected to be small (0.02 mas), and well
below our measured limit over 80 days of 0.10$\pm$0.14 mas (see
\S\ref{sec:obs}).

Proper motion in the cannonball model originates from the superluminal
motion of plasmoids ejected during a supernova explosion with
$\Gamma_\circ\sim$1000 \citep{ddr03}. Dar \& De R\'ujula
(2003)\nocite{dr03} predicted a displacement of 2 mas over the 80 days
of our VLBI experiment assuming plasmoids propagating in a constant
density medium. This estimate was revised downward to 0.55 mas by
incorporating plasmoid interactions with density inhomogeneities at a
distance of $\sim$100 pc within a wind-blown medium \citep{ddr04}.
Neither variant of this model are consistent with our proper motion
limits. A more general problem for the cannonball model is the absence
of rapid fluctuations in the radio light curves of \grb\ 
\citep{bkp+03}. Strong diffractive scintillation is expected between 1
and 5 GHz with a modulation index of order unity and a timescale of a
few hours, because the size of the plasmoids ($\sim$0.01 $\mu$as)
always remain below the Fresnel scale ($\sim$5 $\mu$as) of the
turbulent ionized medium \citep{tc93,wal97}. Strong and persistent
intensity variations in centimeter radio light curves for {\it all}
GRBs are expected in the cannonball model. Strong intensity variations
are not seen for \grb, nor are they expected for the relativistic
blastwave model. Our angular size measurements in \S\ref{sec:size} and
Fig.~\ref{fig:theta} suggest that the expanding fireball is too large
after the first few days to exhibit diffractive scintillation.  There
are moderate variations seen in the radio light curves of \grb\ (25\%
at 4.9 GHz, 15\% at 8.5 GHz and 8\% at 15 GHz) which decrease by a
factor of three from $\sim$3 to 40 d after the burst. Berger et
al.~(2003)\nocite{bkp+03} have attributed this behavior to an
expanding fireball undergoing weak interstellar scintillation and
derive a size for \grb\ of 20 $\mu$as at $\Delta{t}$=15 d.  However,
the details of the change in modulation index depend on knowing
the distance of the screen, the transition
frequency between weak and strong ISS, and the geometry
of the scattering region.  

\section{Conclusions}

We present the first images directly resolving a GRB afterglow 25 and
83 days after the explosion.  The observed expansion velocity of
3-5$c$ can be fit with standard fireball models.  We estimate the
energetics of the burst to have an isotropic equivalent energy, $E_{52}$,
divided by the ambient density of $E_{52}/n_0 \sim 30$ assuming a jet
break at 10 days, and expansion into a constant density circumburst
medium.  Our measurements also place stringent upper limits on the
proper motion of the fireball to less than 0.3 mas over the 80 days
covered, or $<$1.4 mas/year.  These limits are also consistent 
with the standard fireball model.  Much less easy to explain is
the single observation 52 days after the burst of an additional
radio component 0.28 mas northeast of the main afterglow.  This
component requires a high average velocity of $19c$ and cannot
be readily explained by any of the standard models.  Since it is
only seen at a single frequency, it is remotely
possible that this image is an artifact of the calibration.  Other
nearby GRBs would benefit from more frequent time sampling to search
for the presence of similar high-velocity components.

% save space
% Although GRB 030329 has faded considerably, it may still be detectable
% with VLBI techniques, and if so would be more fully resolved 
% with an expected size of 280 mas 1 year after the burst.  
% Fireball models of heating by a single relativistic shock front
% predict that at late times the fireball should look like a ring
% \citep{gps99b}.  To get sufficient signal-to-noise to differentiate
% between Gaussian and ring models would require a significant
% increase in the bandwidth of the observations.

\acknowledgments

We are particularly grateful to the schedulers of the VLBA, GBT,
Effelsberg, WSRT, and Arecibo telescopes for their heroic efforts on
behalf of this program. DAF thanks J. Granot for useful discussions on
the expansion of relativistic jets, and we thank W. Brisken and
S. Chatterjee for checking the results of the proper motion fits.

\begin{deluxetable}{lrrrrrc}
\tabletypesize{\footnotesize}
\tablecolumns{7}
\tablewidth{0pt}
\tablecaption{Observational Summary\label{Observations}}
\tablehead{\colhead{Date}&\colhead{$\Delta t$}&\colhead{Frequency}&\colhead{Integ. Time}&\colhead{BW}&\colhead{Polar.}&\colhead{Instrument}\\
\colhead{}&\colhead{(days)}&\colhead{(GHz)}&\colhead{(min)}&\colhead{(MHz)}&\colhead{}&\colhead{}}
\startdata
01 Apr 2003 & 2.73 & 4.617 & 108 & 16 & 2 & VLBA \\
 & & 4.995 & 108 & 16 & 2 & VLBA \\
 & & 8.421 & 114 & 16 & 2 & VLBA \\
 & & 8.869 & 114 & 16 & 2 & VLBA \\
06 Apr 2003 & 7.71 & 4.617 & 102 & 16 & 2 & VLBA \\
 & & 4.995 & 102 & 16 & 2 & VLBA \\
 & & 8.421 & 96 & 16 & 2  & VLBA \\
 & & 8.869 & 96 & 16 & 2 & VLBA \\
 & & 15.354 & 12 & 32 & 2 & VLBA \\
 & & 22.222 & 6 & 32 & 2 & VLBA \\
22 Apr 2003 & 24.5 & 15.354 & 96 & 32 & 2 & VLBA+EB \\
 & & 22.222 & 104 & 32 & 2 & VLBA+EB \\
19 May 2003 & 51.3 & 15.354 & 90 & 32 & 2 & VLBA+EB+Y27+GBT \\
 & & 22.222 & 118 & 32 & 2 & VLBA+EB+Y27+GBT \\
20 Jun 2003 & 83.3 & 8.409 & 138 & 32 & 2 & VLBA+EB+Y27+WB+AR \\
\enddata
\tablenotetext{*}{NOTE - EB=100m Effelsberg telescope; Y27= phased
  VLA; GBT = 105m Green Bank Telescope; WB = phased Westerbork array;
  305m Arecibo telescope.}
\end{deluxetable}

\clearpage
\begin{deluxetable}{lrrrrrrrc}
\tabletypesize{\footnotesize}
\tablecolumns{9}
\tablewidth{0pt}
\tablecaption{Fitted Parameters\tablenotemark{*}.\label{Big_Gaussian}}
\tablehead{\colhead{Epoch}&\colhead{Freq.}
&\colhead{$S$}&\colhead{rms}&\colhead{Beam}&\colhead{RA$_{\rm offset}$}
&\colhead{DEC$_{\rm offset}$}&\colhead{size}&\colhead{Selfcal?} \\
\colhead{} & \colhead{(GHz)}&\colhead{(mJy)}&
\colhead{(mJy)}&\colhead{(mas)}&\colhead{(mas)}& \colhead{(mas)}&\colhead{(mas)}&\colhead{}\\
\colhead{(1)}&\colhead{(2)}&\colhead{(3)}&\colhead{(4)}&\colhead{(5)}&\colhead{(6)}&\colhead{(7)}&\colhead{(8)}&\colhead{(9)}
}
\startdata
%%%    epoch  frequency  flux rms  ra  dec  size  selfcal?
%%% 2 sigma size limit from (12*sqrt(bmaj*bmin)*rms/peak) 
%%% sigma from  (6 * sqrt(bmaj*bmin)*rms/peak) 
01 Apr 2003 & 4.8 & 3.0 & 0.10 & 2.17 $\times$ 0.99     & $-$0.66 $\pm$ 0.56 & 0.12 $\pm$ 0.56 & $<$0.59 & N \\
  & 8.6 & 8.3 & 0.16 & 1.24 $\times$ 0.57               & 0.0 $\pm$ 0.22 & 0.0 $\pm$ 0.30 & $<$0.19 & N \\
06 Apr 2003 & 4.8 & 5.1 & 0.11 & 2.23 $\times$ 1.05     & $-$0.56 $\pm$ 0.56 & 0.15 $\pm$ 0.56 & $<$0.40 & N \\
  & 8.6 & 15.4 & 0.08 & 1.28 $\times$ 0.58              & 0.07 $\pm$ 0.2 & $-$0.25 $\pm$ 0.2 & $<$0.30 & Y \\
  & 15.4 & 26.8 & 0.39 & 1.16 $\times$ 0.39             & 0.24 $\pm$ 0.2  & $-$0.06 $\pm$ 0.2  & $<$0.12 & Y \\
  & 22.2 & 36.0 & 0.80 & 1.00 $\times$ 0.47             & 0.29 $\pm$ 0.2  & $-$0.26 $\pm$ 0.2  & $<$0.18 & Y \\
22 Apr 2003 & 16.2 & 14.1 & 0.12 & 0.74 $\times$ 0.26   & 0.25 $\pm$ 0.2  & $-$0.35 $\pm$ 0.2  & 0.065 $\pm$ 0.022 & Y \\
  & 22.2 & 11.1 & 0.22 & 0.52 $\times$ 0.18             & 0.26 $\pm$ 0.2  & $-$0.25 $\pm$ 0.2  & 0.077 $\pm$ 0.036 & Y \\
19 May 2003 & 15.4 & 4.0 & 0.08 & 0.67 $\times$ 0.24    & 0.11 $\pm$ 0.2 & $-$0.40 $\pm$ 0.2  & $<$0.10 & Y \\
  & 22.2 & 4.0 & 0.14 & 0.57 $\times$ 0.19              & 0.31 $\pm$ 0.2  & $-$0.03 $\pm$ 0.2 & $<$0.14  & Y \\
20 Jun 2003 & 8.4 & 3.0 & 0.03 & 0.97 $\times$ 0.53     & 0.03 $\pm$ 0.2  & $-$0.24 $\pm$ 0.2 & 0.172 $\pm$ 0.043 & Y \\
\enddata
\tablenotetext{*}{NOTE - Table columns are (1) Epoch of observation;
  (2) observing frequency formed from the average of all IFs; (3) Flux
  density derived from a Gaussian modelfit; (4) naturally weighted
  image rms; (5) uniform weighted beam size; (6) offset in Right
  Ascension and (7) Declination from the April 1st 8.6 GHz position at
  RA 10 44 49.959550, DEC 21 31 17.437881 (J2000); (8) 2$\sigma$ size
  limit or actual FWHM of a circular Gaussian fit to the main
  component; (9) Indication if phase self-calibration was applied.}
\end{deluxetable}

\clearpage

%\begin{figure}
%\plotone{f1.eps}
%\caption{Total intensity image at 15 GHz of \grb\ made with the VLBA,
%GBT and Effelsberg telescopes 51 days after the burst.  The synthesised
%beam is drawn in the lower left corner.  Note the extension of the
%source to the northeast indicating a second component at a 
%separation of 0.28 $\pm$ 0.05 mas (0.80 pc). Contours
%are drawn starting at 0.5 mJy/beam and increase by factors of two.}
%\label{fig:f1}
%\end{figure}

\clearpage

\begin{figure}
\plotone{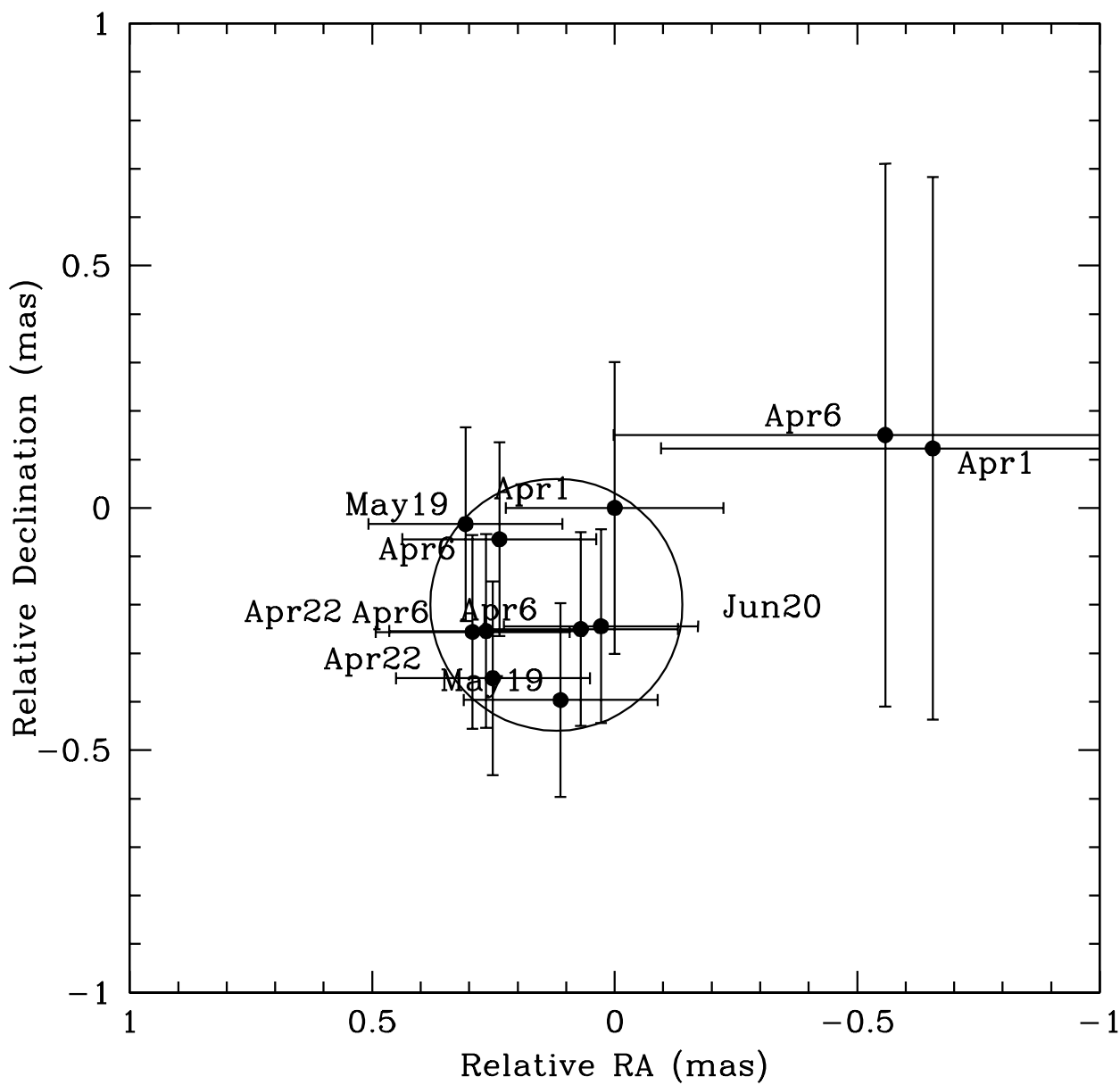}
\caption{The positions derived from the observations in the first five
epochs relative to the first determination on April 1st at 8.4 GHz.
Observations at multiple frequencies at a given epoch have been
plotted separately since they are independent measurements.  A circle
with a radius of 0.26 mas (2$\sigma$) is shown to encompass all
measurements except those taken at 5 GHz, which suffer from systematic
errors (see text for details).  These observations provide a
constraint on the proper motion of 0.10 $\pm$ 0.14 mas over 80 days.}
\label{fig:f1}
\end{figure}

\clearpage

\begin{figure}
\plotone{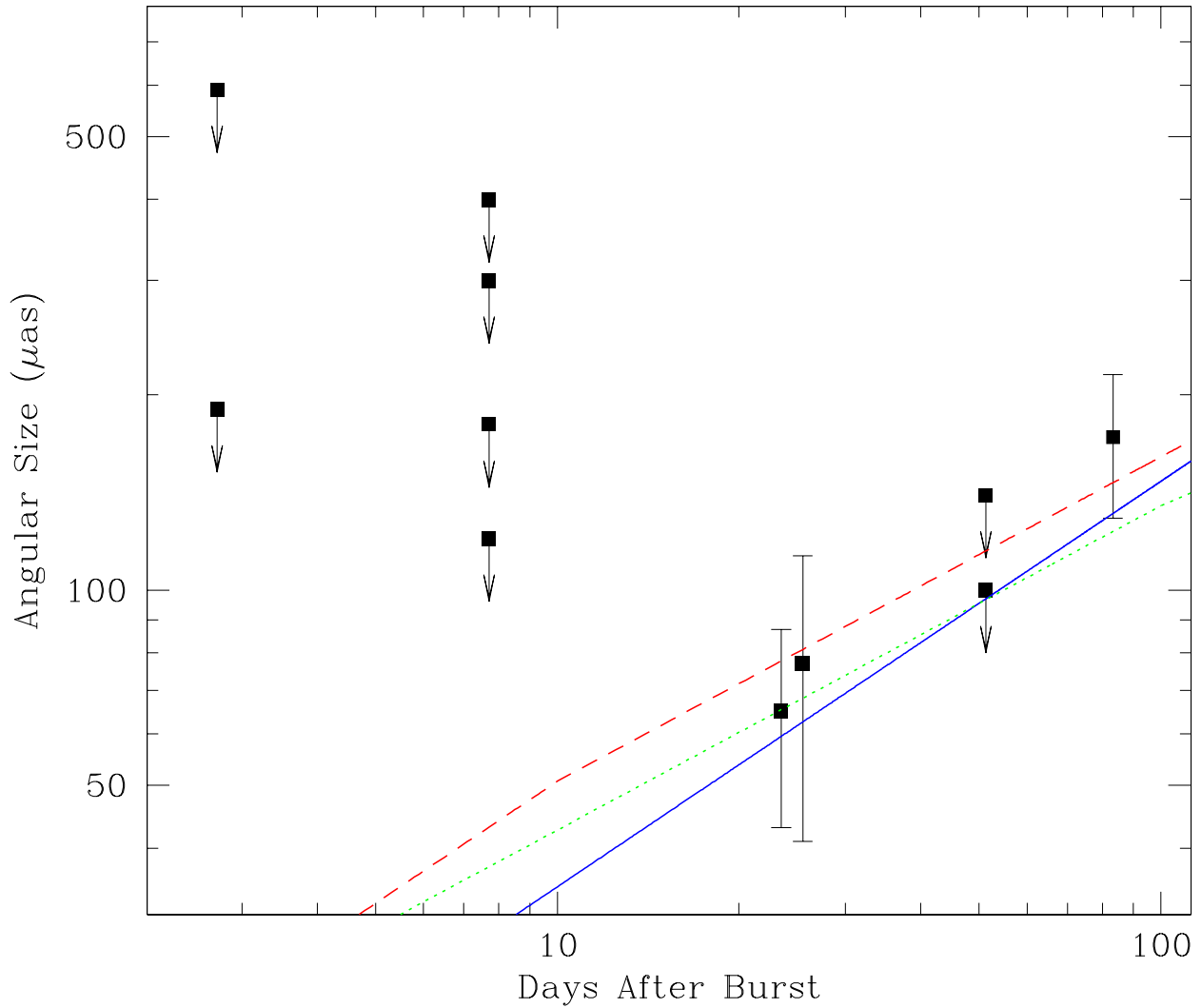}
\caption{Measured angular diameters (or limits) for the radio
afterglow from \grb, along with the expected evolution of the angular
size for different representations of the fireball model. The solid
line is the apparent angular size for a spherical fireball expanding
in a constant density medium with E$_{52}/n_\circ$=1.0. The dotted
line is an early jet model ($t_j$=0.5 d) with E$_{52}/n_\circ$=100,
while the dashed line is a late jet model ($t_j$=10 d) with
E$_{52}/n_\circ$=20. The general tendency is that the more narrowly
collimated the outflow, the larger the energy that is required to
produce agreement with the angular diameter measurements. Note that
the plotted models assume that the circumburst density is constant
(ISM). Wind models over this time range give similar estimates of the
angular diameter.}
\label{fig:theta}
\end{figure}

\clearpage

\end{document}